\documentstyle[preprint,eqsecnum,aps]{revtex}

\begin{document}
\draft
\preprint{CMUMEG-96-007}
\title{Photoproduction evidence for and against hidden-strangeness
states near 2 GeV}

\author{Reinhard A. Schumacher}
\address{Department of Physics, Carnegie Mellon University,
Pittsburgh, PA 15213}
\date{\today}
\author{(Submitted to Phys. Rev. D1)}
\maketitle

\begin{abstract} Experimental evidence from coherent diffractive
proton scattering has been reported for two narrow baryonic
resonances which decay predominantly to strange particles.   These
states, with masses close to 2.0 GeV would, if confirmed, be
candidates for hidden strangeness states with unusual internal
structure.  In this paper we examine the literature on strangeness
photoproduction, to seek additional evidence for or against these
states. We find that one state is not confirmed, while for the other
state there is some mild supporting evidence favoring its existence. 
New experiments are called for, and the expected photoproduction
lineshapes are calculated.
\end{abstract}


\section{INTRODUCTION}

Experimental claims of new, possibly-exotic states must always be
treated with skepticism and caution.  This is especially true when the
structure or quantum numbers of the claimed states are not
established.  A traditional verification test is to seek evidence of
their existence in reaction channels different than those of the
original claim. In this paper I wish to shed additional light on a
pair of fairly narrow baryonic resonances which have been reported
recently by the SPHINX collaboration working at IHEP ~\cite{gol}.  

The two states were seen in coherent diffractive production using 70
GeV protons on a carbon target in the reactions:
\begin{equation}
 p + C \rightarrow [\Sigma^{0} + K^{+}] + C 
\end{equation} and
\begin{equation} p + C \rightarrow [\Sigma^{0}(1385) + K^{+}] + C 
\end{equation} where the brackets signify that the signal was seen in
the invariant mass spectra of these strange particles. Coherent
production off the nucleus was isolated by cutting on low transverse
momentum, $p_{T}$, of the produced particles.  In the first case, for
$p_{T}^{2}<0.1 $ GeV$^{2}$, a structure labeled $X(2000)$ was
identified in the 
$K^{+} \Sigma^{0}$ invariant mass spectrum with 
$M=1996\pm6$ MeV, and $\Gamma=99\pm17$ MeV.  (A more recent and
complete set of results from this group were discussed by L.
Landsberg at the PANIC '96 Conference ~\cite{lan}; here the numbers
were slightly revised to
$M=1996\pm7$ MeV, and $\Gamma=99\pm21$ MeV.)  In the second case, for
$p_{T}^{2}<0.02$ GeV$^{2}$, a structure labeled $X(2050)$ was
identified in the 
$K^{+}\Sigma^{0}(1385)$ invariant mass spectrum with 
$M=2052\pm6$ MeV, and $\Gamma=35^{+22}_{-35}$ MeV.  The apparent width
of the $X(2050)$ state was consistent with the mass resolution of the
experiment.  No signal for these states was seen in the non-coherent
part of the data, i.e. at larger transverse momentum.  No further
analysis to uncover the spin, isospin or angular distributions of
these states was presented.

Both the $X(2000)$ and $X(2050)$ structures are quite narrow for such
massive baryonic resonances, where typical widths are several hundred
MeV.  If these states are confirmed, their narrowness hints at unusual
internal structure.  Furthermore, the states were seen to decay 
predominantly to strange particles, with no positive signal seen in
other final states such as $\Delta\pi$,
$p\pi\pi$, and $\Lambda K$.  Most nucleon resonances have
strange-particle decay branches in the range of a few percent (for
those few cases where these branches have been measured).  Golovkin
{\em et al.}~\cite{gol} quoted lower limits on the order of unity for
branching ratio measurements of the $K\Sigma$ decays to the other
decay channels.  The dominance of the strange-particle decays of
these states was cited as another unusual feature of these states. 
The authors interpreted the states as 5-quark exotics containing
valence
$s-\overline{s}$ quark pairs, and hypothesized that the states may be
examples of
$(qqq)-(q\overline{q})$ structures with color-octet bonds, or
$(qq)-(qq\overline{q})$ with color-sextet bonds ~\cite{dec}. In this
interpretation, the narrowness of the states was due to an angular
momentum barrier between the colored quark clusters which inhibits
decay, and the decoupling from pionic decay channels stemmed from an
OZI suppression of decays that involve annihilation of the strange
quarks.

Another possible interpretation is that these states, if they are
confirmed, are `molecular' states of strange particles.  The
$a_{o}/f_{o}(980)$ states are established examples ~\cite{wei} of
states which find their most natural interpretation as molecular
states, which means that their wavefunctions have large components of
lightly bound
$K\overline{K}$ mesons.  The $\Lambda(1405)$ may have a similar
structure. In the present case, we note that the $X$-state masses are
suggestive of
$K^{*} Y$ bound states, as shown in Table \ref{table1}.  The
$K^{*}(892)$ has a width of 50 MeV, while the `binding energies' of
the
$X$ states observed in Refs ~\cite{gol}~\cite{lan} are in the range
of a few tens of MeV. The states may be analogous to the
$a_{o}/f_{o}(980)$ states or the $\Lambda(1405)$, with the interesting
difference that one of the molecular partners, the $K^{*}$,  is wide
in mass compared to the effective width of the bound state. The $X$
molecular states could therefore decay directly to $K^{*}+Y$ through
the ``tails" of the mass distributions.  Alternatively, the decays
could involve more complicated quark rearrangements, would be
correspondingly slower, and yield particles such as the observed
kaons and Sigmas.  These latter decays would be analogous to the 
fate of the
$\Lambda(1405)$, which decays only to $\Sigma\pi$.

Photoproduction offers a way to confirm the existence of these states
and to test the internal structure hypotheses. A photon in the GeV
energy range can behave as a vector-dominance $\phi$, so
photoproduction is a natural possibility for injecting the right
quark content into the nucleonic system in the $s$ channel via the
reaction
$\gamma + p \rightarrow X \rightarrow K + Y$, as illustrated in
Fig.~\ref{fig1}.  The
$X$ states would appear as
$s$-channel bumps, and would be straightforward to detect using a
tagged bremsstrahlung beam and a suitable spectrometer, provided the
$s$ channel is dominant.  In both the `exotic' and the `molecular'
interpretations of these states, the line shapes for a given channel
should be Breit-Wigner resonances modified due to the opening of the
$K^{*} Y$ channels.  This is discussed in some detail below.  The
integrated branching fractions for
$K Y$ versus
$K^{*} Y'$ final states ought to be strong clues to the structure of
these states. In principle, using photoproduction with good ($\approx
5$ MeV) energy resolution should make it possible to measure the line
shapes directly.

\section{PHOTOPRODUCTION DATA}

The hidden-strangeness states introduced above would be produced as
$s$ channel resonances in photoproduction centered at photon energies,
$E_{\gamma}$, of
\begin{displaymath} X(2000)\rightarrow\Sigma^{0} + K^{+}\hspace{1.0
in}E_{\gamma} = 1700 MeV
\end{displaymath} and
\begin{displaymath} X(2050)\rightarrow\Sigma(1385) + K^{+}\hspace{.6
in}E_{\gamma} = 1750 MeV
\end{displaymath} Bubble chamber experiments published in the late
sixties ~\cite{cbc} ~\cite{abbhhm} provide us with a glimpse of
strange particle photoproduction.  They typically did not achieve
enough resolution or statistics to make detailed analyses of isobar
formation. They did have very good acceptance for all charged final
states, and thus were able to broadly measure and categorize whole
classes of reactions.  The total strangeness photoproduction cross
section from the Cambridge Bubble Chamber Group ~\cite{cbc} shows a
fairly dramatic increase, from 2 to 10 micro-barns, just below 2 GeV,
as seen in Fig.~\ref{fig2}. The increase cannot be explained as the
sum of all available two-body channels ~\cite{ras}, nor has it been
studied in terms of more complex final states.  Thus, no useful
information can be gleaned for our present purpose from the total
cross section.  

In magnetic spectrometer experiments it should be possible to pick out
strong $s$-channel resonances directly, simply by detecting a
$K^{+}$ at any kinematically allowed fixed angle. Feller {\em et al.}
~\cite{fel} used the bremsstrahlung difference method to detect
$\gamma + p \rightarrow \Sigma^{0} + K^{+}$ at approximately fixed $t$
(fixed spectrometer angle) at Bonn.  Fig.~\ref{fig3} shows their data
for $K^{+}$ production with a recoiling $\Sigma^{0}$.  It is
surprising and unfortunate that a data point at an energy
corresponding to the mass of the $X(2000)$ is missing.  The other
data points give no hint of a 100 MeV wide structure is in this
region. Fortunately, an experiment by G\"{o}ing {\em et al.} from
DESY covered a similar range of $W$ (c.m. energy) at larger $t$
~\cite{goi}.  This was a bremsstrahlung experiment in which the
photon endpoint was scanned over the range of interest, and results
extracted from the excitation curves at a few fixed spectrometer
momenta. Fig.~\ref{fig3} shows their results for 
$\Sigma^{0}$ production, which range up to just 2 GeV in mass, where
the centroid of the $X(2000)$ state of should be.  There is no sign
of an
$s$-channel resonance centered at 2 GeV with a 100 MeV width in these
data. Both groups fit their data with phenomenological resonance
models and obtained qualitative agreement with their data sets.  
Thus we can conclude that there is no support in the existing
photoproduction data for the $X(2000)$.  It would perhaps be
interesting to obtain higher statistics samples of this kind of data
to be certain.

Photoproduction of the $\Sigma(1385)$ has not been extracted from any
spectrometer experiment. We are forced to reconsider the sparse bubble
chamber data ~\cite{abbhhm}~\cite{cbc2}, which are collected in
Fig.~\ref{fig4}.  The
$\Sigma(1385)$ decays to $\Lambda \pi$ 88\% of the time.  Thus we
consider the data for 
$\gamma + p  \rightarrow  K + \Lambda + \pi$ as a function of photon
energy.   In the CBCG study~\cite{cbc2}, the $\Lambda \pi$ invariant
mass spectrum (not shown here) summed over all energies had a peak
corresponding to
$\Sigma(1385)$ production; it comprised about 26\% of the $\Lambda +
\pi + K$ final states. Thus, this final state had a significant
component going through the particular two-body decay of interest
here.    Examining Fig.~\ref{fig4}, we find a possible bump at
roughly the right energy ($E_\gamma = 1.75$ GeV) to form the
$X(2050)$.  This bump amounts to no more that one high channel with a
one-sigma error bar, but it is the only ``high" channel in the
spectrum.  Note that the error bars in this plot look too large to be
purely statistical. The number of raw counts in each bin of this
histogram is not clear, but perhaps the significance of the peak is
statistically greater than it appears from the error bars alone. In
any event, this bump is the only hint of narrow
$s$-channel structure in any of the final states from this CBCG
measurement.

ABBHHM published a comparable spectrum from their
experiment~\cite{abbhhm}, as also shown in Fig.~\ref{fig4}.  In this
case, the single specific final state was 
$K^{0} + \Lambda + \pi^{+}$, hence the smaller cross section.  Once
again, the highest channel is near the photon energy corresponding to
formation of an
$X(2050)$, albeit with an enormous statistical uncertainty.  Thus
there are two photoproduction measurements containing a hint of a
feature which could be related to a fairly narrow $s$-channel
resonance near a mass of 2050 MeV. It would be interesting,
therefore, to accumulate some new data in this energy range of
photoproduction to clarify this situation~\cite{jefflab}.  If the
high channels seen in the old experiments are related to an
$X(2050)$, it is possible  that a simple
$s$-channel scan will reveal the state.

\section{LINESHAPES}

Henceforth we consider only an $X(2050)$ state.  If it exists,  and if
it decays into both
$K\Sigma(1385)$ and
$K^{*}\Sigma^{o}$, it would clearly be valuable to measure the
branching ratio for these final states.  This information will be an
important clue to the internal structure of the state.  Because the
state sits close to the
$K^{*}\Sigma^{o}$ threshold, and may be a bound `molecule' of these
particles, measurement of the branching ratio may depend on knowledge
of the distorted lineshapes in these two channels. These lineshapes
can perhaps be measured in photoproduction experiments at, for
example, Jefferson Lab, using a tagged photon beam and a kaon
spectrometer.  One can then ask, how strongly are the lineshapes
distorted due to the proximity of the
$K^{*}\Sigma^{o}$ threshold?

As a simple model calculation we consider the $X(2050)$ decaying to
just the two channels mentioned above.  We compute the lineshapes in
these channels using the Flatt\'{e} formula for a single resonance
decaying to two final states, as discussed in Chung {\em et
al.}~\cite{chung}.  The
$X(2050)$ appears as a ``normal'' resonance in 
$K\Sigma(1385)$ (channel 1), and highly distorted in $K^{*}\Sigma^{o}$
(channel 2).  The mass projection, $P(m)$, of channel 1 is related to
a Lorentz invariant T-matrix element
$\hat{T}_{11}(m)$, and a mass-dependent density of states
$\rho_{1}(m)$, by  
$P(m) \propto | \rho_{1}(m)\widehat{T}_{11} |^{2}$   The T-matrix
element for channel 1 is written as
\begin{equation}
\widehat{T}_{11}(m) = \frac{\gamma_{1}^{2}m_{0}\Gamma_{0}}{ m_{0}^{2}
- m^{2} - i m_{0} \Gamma_{0}(\rho_{1}(m)\gamma_{1}^{2} + \rho_{2}(m)
\gamma_{2}^{2}) }
\end{equation} with an analogous expression for $\widehat{T}_{22}$ of
channel 2. The `reduced' widths for the two channels, denoted
$\gamma^{2}_{1}$ and 
$\gamma^{2}_{2}$, satisfy  $\gamma^{2}_{1} + \gamma^{2}_{2} = 1$.  The
parameters $m_{0}$ and $\Gamma_{0}$ are the mass and width one would
estimate if channel 1 were due to decay of a single isolated
resonance. The observed mass,
$m_{X}$, and the observed (though perhaps distorted) width,
$\Gamma_{X}$, seen in channel 1 are related to $m_{0}$ and
$\Gamma_{0}$ by
\begin{equation} m_{0}^{2} = m_{X}^{2} +
m_{X}\Gamma_{X}\frac{\rho_{2}(m_{X})}{\rho_{1}(m_{X})}
\frac{\gamma_{2}^{2}}{\gamma_{1}^{2}}
\end{equation} and
\begin{equation}
\Gamma_{0} = \Gamma_{X}\frac{m_{X}}{m_{0}} \frac{1}{\rho_{1}(m_{X})}
\frac{1}{\gamma_{1}^{2}}.
\end{equation} The density of states factors $\rho_{1}$ and 
$\rho_{2}$ are related to the available center-of-mass momentum, $q$,
when a state of mass $m$ decays to two final states masses $m_{a}$
and $m_{b}$.  The standard form ~\cite{chung} when $m_{a}$ and
$m_{b}$ have negligible widths is
$\rho(m,m_{a},m_{b}) = 2q(m,m_{a},m_{b}) / m $, which approaches unity
as $m \rightarrow \infty$ and vanishes below the mass threshold. If
either
$m_{a}$ or
$m_{b}$ are in turn states of finite width, such as the $K^{*}$ or
$\Sigma(1385)$, then the density of available states rises with mass
more gradually than an abrupt threshold.  For the present model
calculation we folded  
$\rho(m,m_{a},m_{b})$ with an s-wave Breit-Wigner lineshape
$P_{i}(m_{i})$, $i=a,b$, for the ``broad'' final state particle.  For
the density of states in channel 2 we used
\begin{equation}
\rho_{2}(m) = \int_{0}^{m - m_{\Sigma}}
\rho(m,m_{K^{*}},m_{\Sigma^{0}}) P_{K^{*}}(m_{K^{*}}) dm_{K^{*}} 
\end{equation} The convolved form for $\rho_{2}$ is intended to
account for the width of the
$K^{*}$ final state, and the fact that not all of the phase space for
channel 2 is available at a given value of the mass of the state
$m_{X}$. 
$P_{K^{*}}(m_{K^{*}})$ is a simple Breit-Wigner lineshape for the
$K^{*}$.

In Fig.~\ref{fig5}, the results of this calculation are shown for a
series of possible widths observed in channel 1, the `normal'
resonance channel.  The ratio of reduced widths,   $\gamma^{2}_{1} /
\gamma^{2}_{2}$, was set equal to unity because the lineshapes were
found to be not drastically sensitive to this ratio for values above
0.1.  The main sensitivity of the lineshape was seen in the dependence
on width,
$\Gamma_{X}$, in the  $K^{*}\Sigma^{o}$ case.  The solid lines are for
$\Gamma_{X} = 35 MeV$, the nominal observed width reported in the
experiment~\cite{gol}.  For this width, the second channel is
predicted to have a broad structure with a second maximum in the
mass.  The shape of the
$K^{*}(892)\Sigma^{o}$ distribution varies rapidly as the observed
width of the
$K\Sigma(1385)$ channel is varied from 20 to 100 MeV.  On the other
hand, the shape of the
$K\Sigma(1385)$ channel is rather $in$sensitive to the opening of the
$K^{*}\Sigma^{o}$ channel. Thus, careful measurement of the
$K^{*}\Sigma^{o}$ final state lineshape would help define the
$X(2050)$ width. No strong variations in lineshape was found when the
position of the mass centroid was varied over the roughly 20 MeV
range of experimental uncertainty.   

From this study it may be concluded that careful measurements of the
lineshapes of the $X(2050)$, if it exists, would not be sensitive to
the ratio of reduced widths.  However, the $K^{*}(892)\Sigma^{o}$
final state is quite sensitive to the precise width of the
$X(2050)$.  

\section{CONCLUSIONS}

We have shown that strangeness photoproduction data off the proton 
covering the
$s$-channel mass range around 2000 MeV do not confirm a 100 MeV wide
state seen at this mass in coherent diffractive proton scattering
~\cite{gol}.  On the other hand, strangeness photoproduction data
near a mass of 2050 MeV mildly supports the existence of another
state seen in the same proton scattering experiment.  If confirmed by
new and better experiments in photoproduction, either of these states
may be of considerable interest as exotic ``hidden strangeness''
states, since their decays are dominantly to strange particles.  A
computation of the distorted lineshapes of the $X(2050)$ state shows
that it will appear nearly as a normal resonance in the
$K\Sigma(1385)$ channel, and strongly distorted in the
$K^{*}(892)\Sigma^{0}$ channel.   Experiments at the new generation
of photon facilities should make such experiments feasible; initial
experiments would be straightforward `single-arm' kaon measurements
using a tagged photon beam.

\section*{ACKNOWLEDGMENTS}

I thank Zhen-Ping Li and Curtis Meyer for helpful discussions, as well
as L.G. Landsberg for discussions of the data.  This work was
supported by DOE contract DE-FG02-87ER40315.

\begin{figure}
\caption{$s$-channel photoproduction of strangeness-rich intermediate
states via vector dominance $\phi$'s.}
\label{fig1}
\end{figure}

\begin{figure}
\caption{Total cross section for strange particle photoproduction. The
data are from Ref. \protect\cite{cbc}.}
\label{fig2}
\end{figure}

\begin{figure}
\caption{Exclusive strangeness photoproduction of the $\Sigma^{0}$. 
Data are from Refs.\protect\cite{fel} and \protect\cite{goi}, and
include points from other experiments.  Location expected for
$s$-channel production of an
$X(2000)$ with a width of 99 MeV is indicated by the arrow and the
dashed lines.}
\label{fig3}
\end{figure}

\begin{figure}
\caption{Total cross sections for all charge combinations for $\gamma
+ p \rightarrow \{\Lambda, \Sigma\} + K + \pi$ from Ref.
\protect\cite{cbc2} and for
$\Lambda + K^{0} + \pi^{+}$ from Ref.\protect\cite{abbhhm}.  Location
expected for  $s$-channel production of an $X(2050)$ with a width of
35 MeV is indicated by the arrow and the dashed lines.}
\label{fig4}
\end{figure}

\begin{figure}
\caption{Predicted lineshapes of the $X(2050)$ decaying to two final
states for equal reduced widths $\gamma^{2}_{i}$.  The observed FWHMs
of the
$X(2050)$ are the nominal 35 MeV (solid lines), 20 MeV (short dash),
50 MeV (dotted), and 100 MeV (dot-dash).  In each case the curves are
arbitrarily normalized.}
\label{fig5}
\end{figure}

\begin{table}
\caption{Mass comparisons of `narrow' states with `molecular'
combinations of known mesons and baryons.}
\begin{tabular}{cccc} `Molecular'        & Constituent &
Observed          &`Binding\\ Structure          & Mass Sum    &
State             &Energy'\\
                   &(MeV)        &                   &(MeV)\\
\hline\hline
$K^{+} K^{-}$          &  988    &  $a_{0}/f_{0}(980)$     & 8 \\

$K N$                  &  1435   &  $\Lambda(1405)$  & 27 \\

$K^{*}(892)\Lambda$    &  2007   &  $X(2000)$        &  10 \\

$K^{*}(892)\Sigma^{0}$ &  2084   &  $X(2050)$        &  35 \\

\end{tabular}
\label{table1}
\end{table}


\begin{references}
\bibitem{jefflab} These data can be accumulated at the new generation
of electron laboratories:  Jefferson Lab, Bonn, and Grenoble.  At
Jefferson Lab the data set is planned to come from CLAS experiment
89-004.
\end{references}
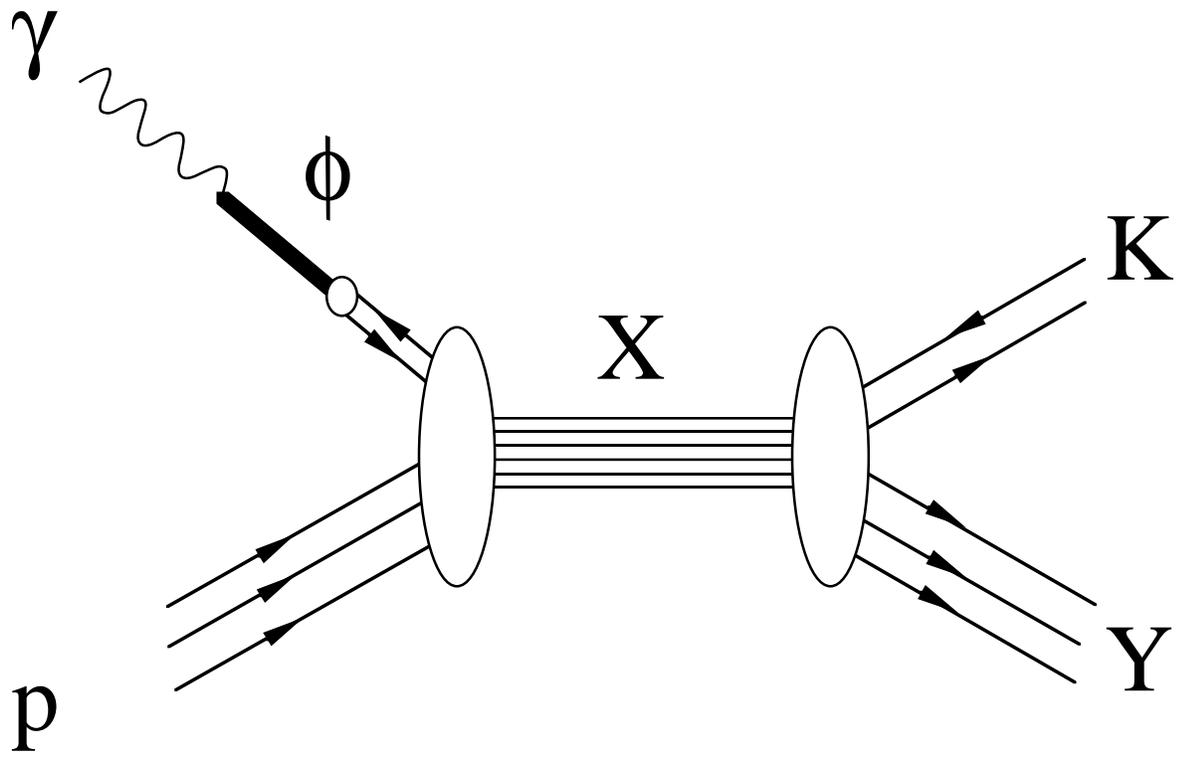
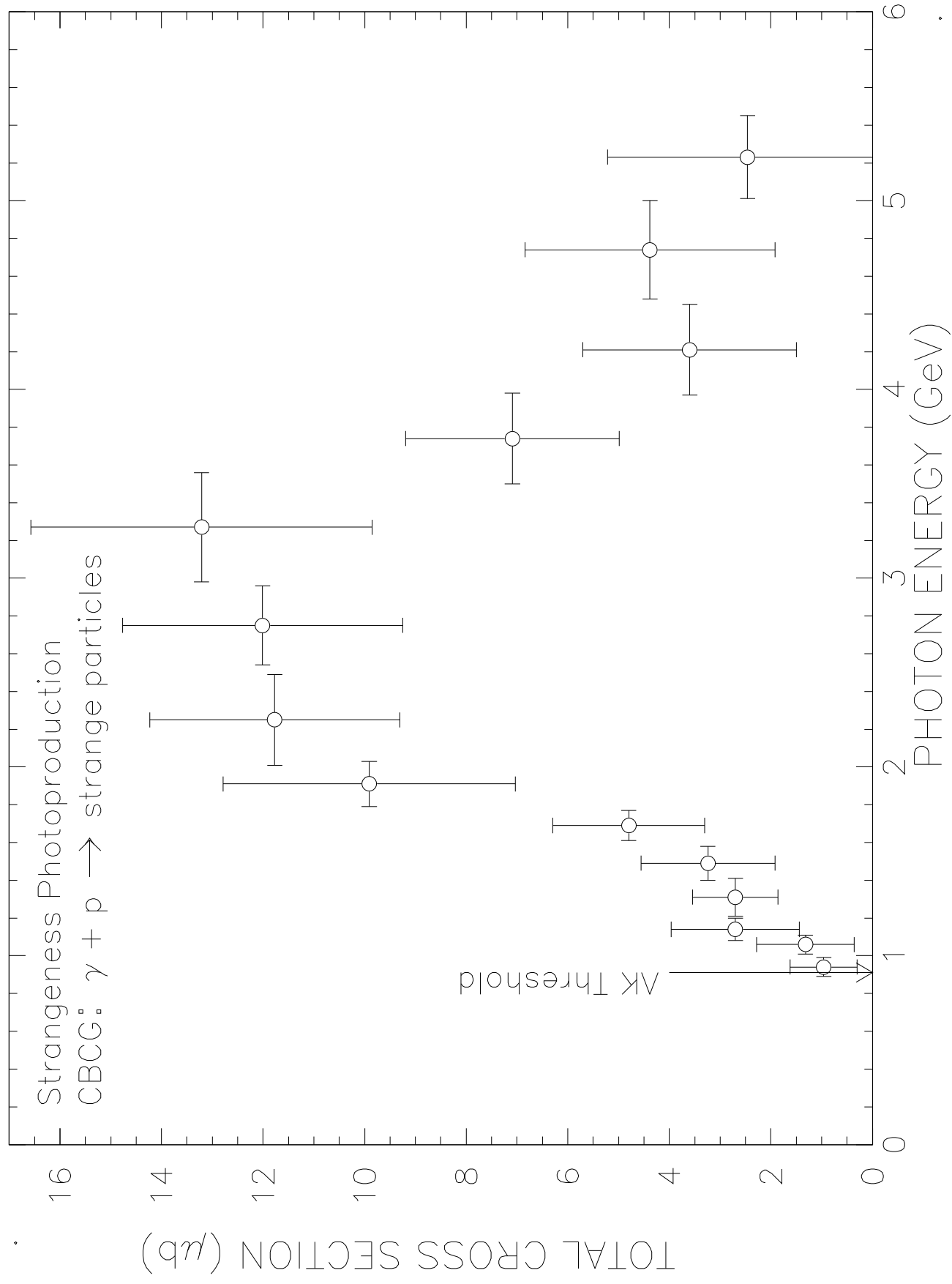
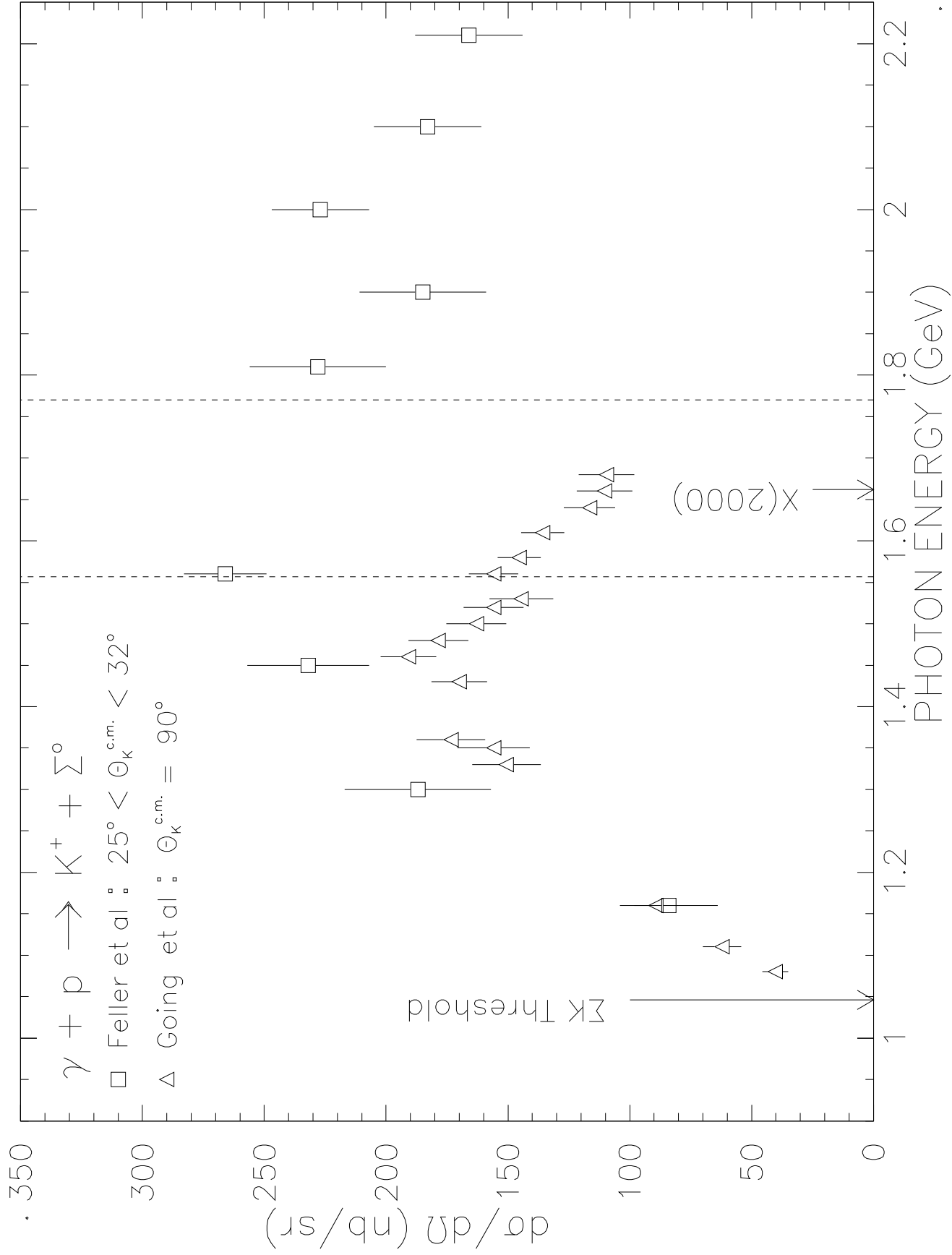
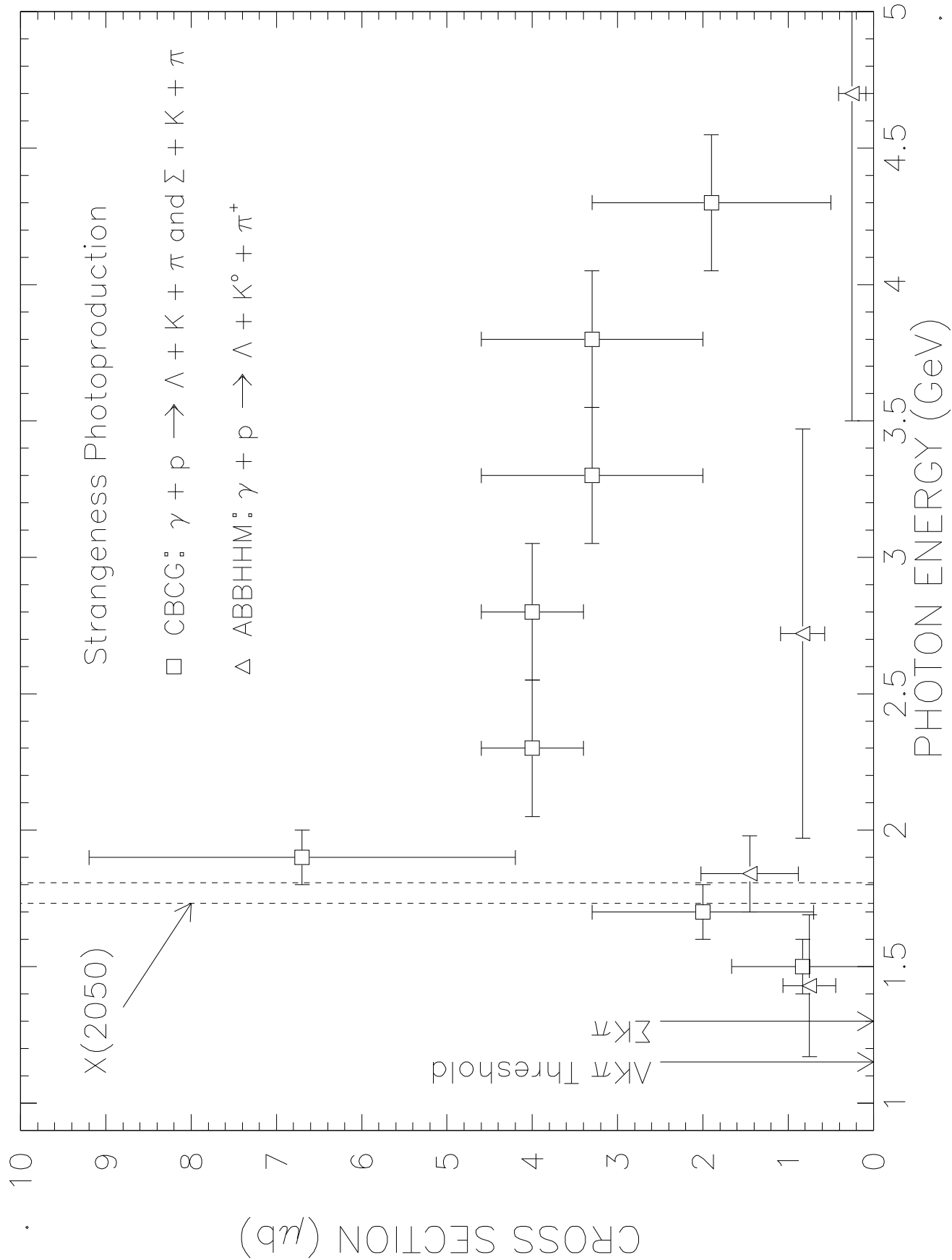
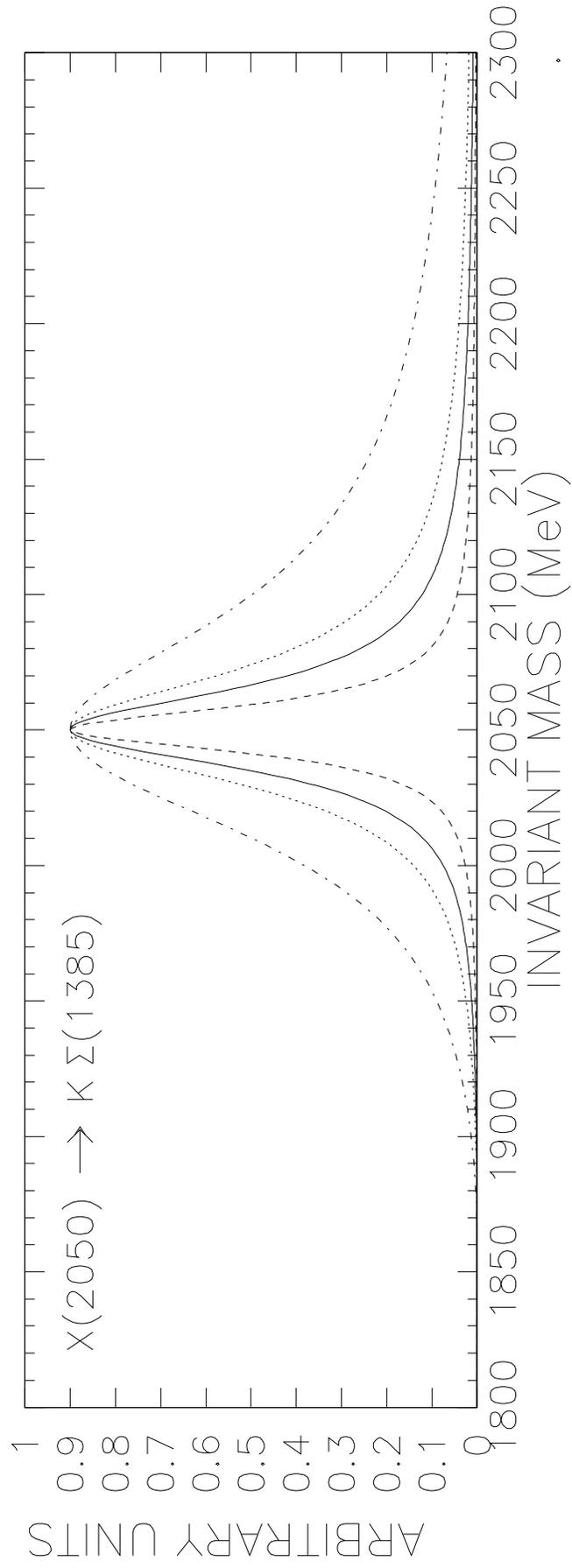
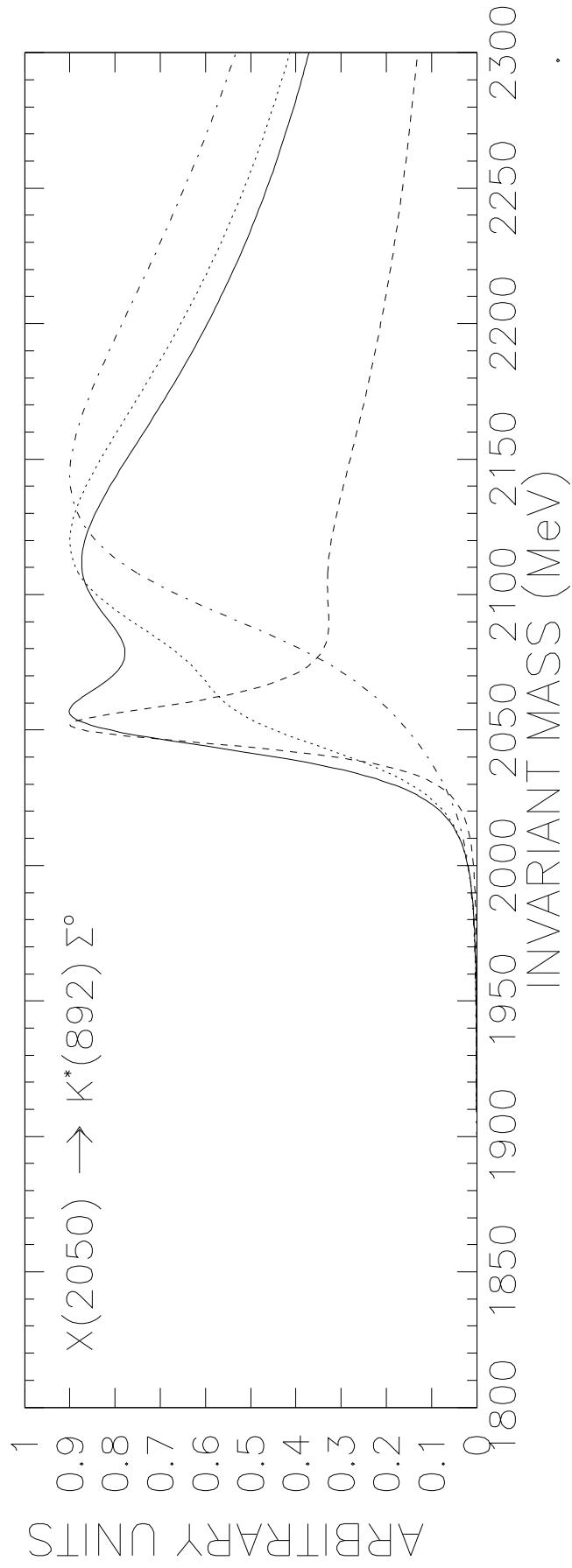
\end{document}